\documentclass[iop]{emulateapj}
\usepackage{amssymb,multirow,color}

\renewcommand{\vec}[1]{\mathbf{#1}}

\renewcommand{\(}{\left(}
\renewcommand{\)}{\right)}

\begin{document}
\title{Residual Energy in Magnetohydrodynamic Turbulence} 
\author{Yuxuan Wang$^1$, Stanislav Boldyrev$^{1,2}$, Jean Carlos Perez$^{1,3}$}
\affil{${~}^1$Department of Physics, University of Wisconsin, Madison, WI 53706, USA\\
${~}^2$Kavli Institute for Theoretical Physics, University of California, Santa Barbara, CA 93106, USA\\
${~}^3$Space Science Center and Department of Physics, University of New Hampshire, Durham, NH 03824, USA}
\date{July 21, 2011}

\begin{abstract}
There is mounting evidence in solar wind observations and in numerical simulations that kinetic and magnetic energies are not in equipartition in magnetohydrodynamic turbulence. The origin of their mismatch, the residual energy $E_r=E_v-E_b$, is not well understood. In the present work this effect is studied analytically in the regime of weak magnetohydrodynamic turbulence. We find that residual energy is spontaneously generated by turbulent dynamics, and it has a negative sign, in good agreement with the observations. We obtain that the residual energy condenses around $k_\|=0$ with its  $k_\|$-spectrum broadening linearly with $k_\perp$, where $k_\|$ and $k_\perp$ are the wavenumbers parallel and perpendicular to the background magnetic field, and the field-perpendicular  spectrum of the residual energy has the scaling $E_r(k_\perp)\propto k_\perp^{-1}$ in the inertial interval. These results are found to be in agreement with numerical simulations. We propose that residual energy plays a fundamental role in Alfv\'enic turbulence and it should be taken into account for correct interpretation of observational and numerical data.   
\end{abstract}

\maketitle

\section{Introduction}
Turbulence plays an important role in a variety of astrophysical phenomena, including amplification of magnetic fields in planetary and stellar interiors, magnetization and angular momentum transport in accretion disks, scattering of cosmic rays, formation of small-scale density structures in the interstellar medium, heating of the solar corona and the solar wind. Theoretical description of magnetized  plasma turbulence is a complicated task.  A valuable guidance for phenomenological modeling is provided by high-resolution numerical simulations \citep[e.g.,][]{brandenburg_n11,kritsuk_etal11}. On the analytical side, a significant progress is made in the limit of weak turbulence, that is, turbulence consisting of weakly interacting linear waves \citep[e.g.,][]{newell_nb01,galtier_nnp00,galtier_nnp02}, which provides a testbed for fundamental ideas in the theory of turbulence, such as scale invariance, locality of interactions, energy cascades, and  anisotropy. Weak MHD turbulence is also interesting on its own as it may play a role in the interstellar medium, in the solar corona and the solar wind, in planetary and stellar  magnetospheres   \citep[e.g.,][]{bhattacharjee_n01,saur_etal02,melrose06,rappazzo_etal2008,chandran10,chandran_etal10}. 

In the present Letter we address the problem of mismatch between kinetic and magnetic energies recently reported in observations of the solar wind turbulence. Our interest is also motivated by a similar mismatch found in recent numerical simulations of magnetohydrodynamic (MHD) turbulence, suggesting that the phenomenon may, in fact, have a fundamental nature rather than reflect a possible non-universality of solar wind turbulence. By employing the framework of weak MHD turbulence we present the first direct analytical derivation of residual energy generation in MHD turbulence. We demonstrate that kinetic-magnetic equipartition gets spontaneously broken by interacting random Alfv\'en waves, even if it is present  initially. Our results indicate that this effect plays a fundamental role in the energy cascade, and it should be taken into account in phenomenological modeling of MHD turbulence. 

For our analysis we write the  incompressible MHD equations in terms of the Elsasser variables:
\begin{equation}
  \(\frac{\partial}{\partial t}\mp\vec v_A\cdot\nabla\)\vec
  z^\pm+\left(\vec z^\mp\cdot\nabla\right)\vec z^\pm = -\nabla P,
  \label{mhd-elsasser}
\end{equation}
where the Elsasser variables are defined as $\vec z^\pm=\vec
v\pm\vec b$. $\vec v$ is the fluctuating plasma velocity, $\vec b$ is
the fluctuating magnetic field normalized by $\sqrt{4 \pi \rho_0}$,
${\bf v}_A={\bf B}_0/\sqrt{4\pi \rho_0}$ is the Alfv\'en velocity corresponding to the
uniform magnetic field ${\bf B}_0$, $P=(p/\rho_0+b^2/2)$ includes the
plasma pressure $p$ and the magnetic pressure, $\rho_0$ is the constant 
mass density, and we neglected the terms describing viscous and resistive 
dissipation. 
One observes that when $\vec z^\mp(\vec x,t)\equiv 0$, an arbitrary
function $\vec z^\pm(\vec x,t)=F^\pm(\vec x\pm\vec v_At)$ is an exact
solution of~(\ref{mhd-elsasser}), which represents a non-dispersive Alfv\'en  wave propagating parallel or anti-parallel to ${\bf B}_0$ with the Alfv\'en speed. Nonlinear interactions are the result 
of collisions between counter-propagating Alfv\'en wave packets.  This qualitative picture plays a central role in phenomenological models of MHD turbulence; comprehensive reviews of recent analytical and numerical results can be found in, e.g., \citep{galtier09,sridhar10,mininni11,newell_r11,tobias_cb11}. 

The ideal MHD system~(\ref{mhd-elsasser}) conserves the two independent Elsasser energies, $E^\pm=\langle |\vec z^\pm|^2\rangle /4$, related to the total energy and cross-helicity, $E=E^++E^-$ and $H_c=E^+-E^-$, respectively.  
In a turbulent state, when energy is supplied to the system at large scales, 
both energies $E^\pm$ cascade toward small scales where they are converted into heat by  viscosity and resistivity. Theories and numerical simulations of MHD turbulence address the Fourier 
spectra of the energies $E^{\pm}(k)$ in the inertial range of scales, that is, scales 
much smaller than the forcing scales and much larger than the dissipation scales. 

In the picture of counter-propagating Alfv\'en modes it is often assumed that energy spectra are the same for both kinetic and magnetic fluctuations, since such energy equipartition holds for individual Alfv\'en wave packets. However, recent 
solar wind observations demonstrate contradictions with this assumption: the Fourier energy spectra of magnetic and velocity fluctuations appear to have different slopes in the inertial interval  \citep{podesta_rg07,salem_etal09,tessein_etal09,chen_etal10,wicks_etal11,li_etal11,chen_etal11}. This is somewhat similar to the mismatch in the spectra found in numerical simulations of strong MHD turbulence \citep{muller_g05,boldyrev_pbp11}. Most intriguing, however, is the fact that significant difference between magnetic and kinetic energies was detected in numerical studies of {\em weak} MHD turbulence \citep{boldyrev_p09} consisting of mostly independent Alfv\'en waves for which equipartition between magnetic and kinetic energies holds exactly. 

The mismatch between magnetic and kinetic energies, the so-called residual energy $E_r({\bf k})=E_v({\bf k})-E_b({\bf k})$, was noted in previous works, e.g.,  \cite[][]{grappin_pl83,zank_ms96,muller_g05}, however, it has not been fully explored in models of MHD turbulence, since such models are often based on the conserved quantities, $E^\pm$, $E$, $H_c$, which do not contain information about the residual energy.  
In this work we present the analytic derivation of the residual energy in the case of weak MHD turbulence. We demonstrate that residual energy is spontaneously generated by interacting Alfven waves even if it is absent initially, it plays a fundamental role in the turbulent dynamics, and it provides a qualitative explanation for the breakdown of magnetic and kinetic equipartition observed in the solar wind. 

\section{The origin of the residual energy}
We concentrate on turbulence of shear-Alfv\'en waves whose polarizations are normal to the guide field, ${\bf z}^{\pm}\perp {B}_0$. We  assume that the guide field is strong;  its strength with respect to the magnetic and velocity fluctuations is characterized by a small parameter $\epsilon\sim  z^\pm/v_A \ll 1$. We also assume that the turbulence is {\em weak}, that is, the wave spectrum in the field-parallel direction is broad enough to ensure that the nonlinear terms are small,  
$k_\|v_A \gg k_\perp z^\pm$,
\label{weak}
where $k_\|$ and $k_\perp$ are typical wave numbers characterizing the spectral width in the field-parallel and field-perpendicular directions and $z^\pm$ are typical (rms) values of the fluctuations. The Elsasser Fourier energies are given by the correlation functions 
\begin{eqnarray}
\langle {\bf z}^\pm({\bf k}, t)\cdot {\bf z}^\pm({\bf k}', t)\rangle=\epsilon^2 v_A^2 q^\pm(k_\|, k_\perp,t)\delta({\bf k}+{\bf k}'), 
\end{eqnarray}
while the residual energy is related to 
\begin{eqnarray}
\langle {\bf z}^+({\bf k}, t)\cdot {\bf z}^-({\bf k}', t)\rangle=\epsilon^2 v_A^2 Q^0(k_\|,k_\perp;t)\delta({\bf k}+{\bf k}'). 
\end{eqnarray}
In these formulas we assume spatial homogeneity and average over a statistical ensemble. The Elsasser energies are real, while the residual correlation function satisfies  $Q^0(k_\|,k_\perp)^*=Q^0(-k_\|,k_\perp)$. Its real part is then the residual energy $e_r(k)={\rm Re} Q^0$. In practical applications one also uses the phase-volume compensated residual energy  defined as 
$E_r(k)=\epsilon^2 v_A^2 e_r(k) 2\pi k_\perp $.  
In previous treatments  \citep{galtier_nnp00,galtier_nnp02}, the residual energy was assumed to be zero as it would be the case for independent $z^+$ and $z^- $ waves. In the present study we do not make such an  assumption, rather, we {\em derive} the equation for the residual energy. 

The derivation is performed perturbatively in~$\epsilon$, following the standard procedure \citep[e.g.,][]{galtier_nnp02}. When the nonlinear terms in Eq.~(\ref{mhd-elsasser}) are neglected, both ${\bf z}^\pm({\bf k})$ oscillate in time and so does the residual energy, 
$Q^0(k,t)=q^0(k)\exp(2ik_\|v_At)$.
When the nonlinear interaction is taken into account, $q^0$ acquires small corrections that one finds, to the first nonvanishing order,  
by integrating the equation:  
\begin{eqnarray}
& \partial_t q^0(k,t)= 
-{\rm e}^{-2ik_\|v_At}\epsilon^2 \int R_{k,pq} \{ q^{+}({\bf q})[q^{-}({\bf p})-q^-({\bf k})] \nonumber  \\
&+q^{-}({\bf q})[q^{+}({\bf p})-q^+({\bf k})] \}{\tilde \delta}(q_\|) \delta({\bf k}-{\bf p}-{\bf q})d^3p\,d^3q.  
\label{bp-eq}
\end{eqnarray}
where $R_{k,pq}=(\pi v_A/2)({\bf k}_{\perp}\times {\bf q}_{\perp})^2({\bf k}_{\perp}\cdot{\bf p}_{\perp})({\bf k}_{\perp}\cdot{\bf q}_{\perp})/(k_{\perp}^2p_{\perp}^2q_{\perp}^2)$, and we introduce a ``smeared'' delta-function ${\tilde \delta}(k_\|)=(1-\exp(-2ik_\|v_A t ))/(i\pi k_\|)$, which approaches  $\delta(k_\|)+(i/\pi){\cal P}(1/k_\|)$ at large $t$. Since we are interested in spontaneous generation of the residual energy, we assume that it is absent initially, 
therefore, in the rhs of Eq.~(\ref{bp-eq}) we neglected the terms proportional to $q^0$. The derivation of Eq.~(\ref{bp-eq}) is standard although lengthy, and it will be presented elsewhere. In what follows we discuss the main consequences of the obtained result.

\section{The spectrum of the residual energy}
First, we notice that the integral in the rhs of Eq.~(\ref{bp-eq}) is essentially nonzero. It therefore describes the {\em generation} of the residual energy by Alfv\'en waves. The important conclusion is that the  residual energy is generated by nonlinear interactions even if it is zero initially. Moreover,  one can directly verify that for all the spectra of interest, the integral in the rhs of Eq.~(\ref{bp-eq}) is positive. As we shall see below, this means that the generated residual energy is negative, that is, the magnetic energy exceeds the kinetic one.\footnote{This result is a statistical effect of many independent  collisions of Alfv\'en waves. It does not imply that a particular collision of two Alfv\'en wave packets necessarily produces negative residual energy.} 

A simple evaluation of the integral in Eq.~(\ref{bp-eq}) can be performed for the case of balanced MHD turbulence, that is, for $q^+=q^-$. In this case, the Elsasser energy spectra have the well known scaling \citep[e.g.,][]{ng_b96,galtier_nnp00}: $q^+ = q^- =  f(k_\|)k_\perp^{-3}$, where $f(k_\|)$ is arbitrary, and we assume that it is smooth at least initially. For the reasons that will be clear momentarily, we estimate the integral in the rhs of Eq.~(\ref{bp-eq}) at $k_\|=0$: 
\begin{eqnarray}
\int R_{k,pq}\{\dots \}{\tilde \delta}(q_\|) \delta({\bf k}-{\bf p}-{\bf q})d^3p\,d^3q =\alpha k_\perp^{-2} v_A,
\label{alpha}
\end{eqnarray}
where the numerical factor is $\alpha \sim f^2(0)$.\footnote{A more precise numerical integration gives $\alpha \approx 0.065 \pi f^2(0)$, however, our discussion of scaling properties of the residual energy does not require the knowledge of numerical coefficients with such precision.}  Non-vanishing of this term at $k_\|=0$ implies that Eq.~(\ref{bp-eq}) has a solution diverging in time, which is unphysical. The reason is that when $q^0$ becomes large one cannot neglect the nonlinear terms containing $q^0$ in Eq.~(\ref{bp-eq}). As we shall see, in order to obtain the convergent result, it is enough to retain the terms linear in $q^0$, which are proportional to $ q^\pm q^0$. Those terms describe relaxation of the residual energy due to its interaction with the Elsasser energies. Their structure can be found from a simple estimate, if   one notices that these terms should have the same dimensional structure as the rhs term in Eq.~(\ref{bp-eq}), if one replaces one of $q^\pm$ by $q^0$. The corresponding relaxation term can thus be added to the rhs of Eq.~(\ref{bp-eq}) in the form $\gamma q^0$, with the relaxation rate $\gamma \sim -\epsilon^2 v_A k_\perp^4 q^{\pm}(k)$. The estimate of the relaxation rate at $k_\|=0$ gives: 
\begin{eqnarray}
\gamma(0, k_\perp)=-\beta \epsilon^2 k_\perp v_A,
\label{gamma} 
\end{eqnarray}
where $\beta\sim f(0)$.\footnote{It is easy to see that $\gamma$ coincides with the inverse time of nonlinear spectral energy transfer in weak MHD turbulence, \citep[e.g.,][]{ng_b96,galtier_nnp00}.} 

Collecting the results (\ref{alpha}) and (\ref{gamma}), we rewrite Eq.~(\ref{bp-eq}) in the form:
\begin{eqnarray}
\partial_t q^0(k,t)=-\epsilon^2 \beta k_\perp v_A q^0(k,t)-\epsilon^2 \alpha k_\perp^{-2}v_A e^{-2ik_\|v_At},
\label{model}
\end{eqnarray}
which has to be solved together with the initial condition $q^0=0$. The solution is given by 
\begin{eqnarray}
q^0(k,t)=-\frac{\alpha \epsilon^2 k_\perp^{-2}}{\beta \epsilon^2 k_\perp -2ik_\|}e^{-2ik_\|v_At},
\end{eqnarray}
and the residual energy is found as
\begin{eqnarray}
e_r(k_\|,k_\perp)={\rm Re}Q^0=-\frac{\alpha \beta \epsilon^4 k_\perp^{-1}}{\beta^2 \epsilon^4 k_\perp^2 +4k_\|^2}.
\label{er}
\end{eqnarray}
The residual energy can be re-written as
\begin{eqnarray}
e_r(k_\|,k_\perp)=-\alpha \pi \epsilon^2 k_\perp^{-2}\Delta(2k_\|), 
\label{res}
\end{eqnarray}
where the $\Delta$ function has support inside a narrow wedge-shaped domain: 
\begin{eqnarray}
k_\|\leq \beta \epsilon^2 k_\perp/2. 
\label{wedge} 
\end{eqnarray}
According to the result $\lim_{\epsilon \to 0}\epsilon/(x^2+\epsilon^2)=\pi \delta(x)$, the function $\Delta$ would turn into the $\delta$-function in the limit~$\epsilon\to 0$. This explains why the region around $k_\|=0$ was relevant in our estimates (\ref{alpha}) and (\ref{gamma}). We will refer to such a spectrum as ``condensate''. It is also convenient to define the field-perpendicular spectrum of the residual energy,
\begin{eqnarray}
E_r(k_\perp)= \int \epsilon^2 v_A^2 e_r(k_\|, k_\perp)2\pi k_\perp d k_\| =-\alpha \pi^2 \epsilon^4 v^{2}_Ak_\perp^{-1}.
\label{erperp} 
\end{eqnarray}
Expressions (\ref{er}-\ref{erperp}) are the main results of the present work.


\section{Numerical results}
In this section we test our analytic predictions in numerical simulations. 
In the presence of a strong guide field, the universal properties of MHD turbulence can be studied  in the framework of Reduced MHD,  
which can be effectively used in numerical simulations of both strong and weak turbulence, e.g., \citep{dmitruk_etal2003, oughton_etal2004,galtier_c06,perez_b08,rappazzo_etal10}. In our numerical set-up, the $\vec z^+$ and $\vec z^-$ fields are {\em independently} driven by Gaussian distributed random, short-time correlated forces with zero mean and prescribed variances, so that no residual energy is supplied by the driving routine. The forces are applied at $k_{\|} =1,\dots ,16$ \& $k_\perp=1,2$ in the Fourier space. The box is elongated in the field-parallel direction according to $L_\|/L_\perp =B_0/b_{rms}$.  We employ a fully dealiased pseudo-spectral method;  the run has the resolution of $512^3$, the Reynolds number $Re=2400$, and it is averaged over 41 large-scale eddy turnover times. We choose viscosity equal to magnetic diffusivity. The details on the code and the numerical set-up can be found in \citep{perez_b10}. 

\begin{figure}[htbp]
\includegraphics[width=\columnwidth]{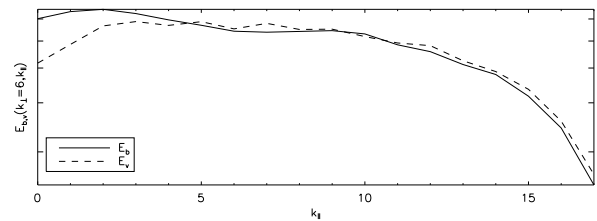}
\caption{The field-parallel spectra of magnetic (solid line) and kinetic (dashed line) energies  at $k_\perp=6$ for weak balanced 
MHD turbulence. The magnetic energy exceeds the kinetic one at small $k_\|$. Note the logarithmic scale on the vertical axis.}
\label{fig:cond}
\end{figure}

\begin{figure}[htbp]
\includegraphics[width=\columnwidth]{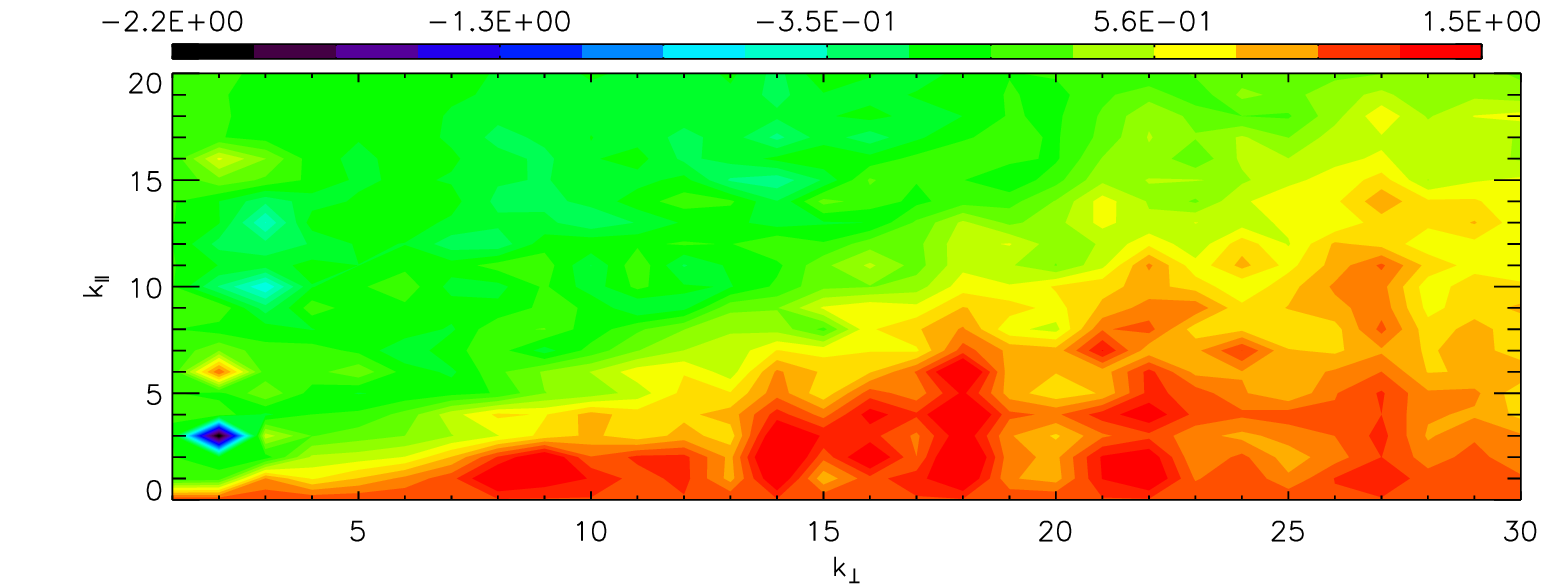}
\includegraphics[width=\columnwidth]{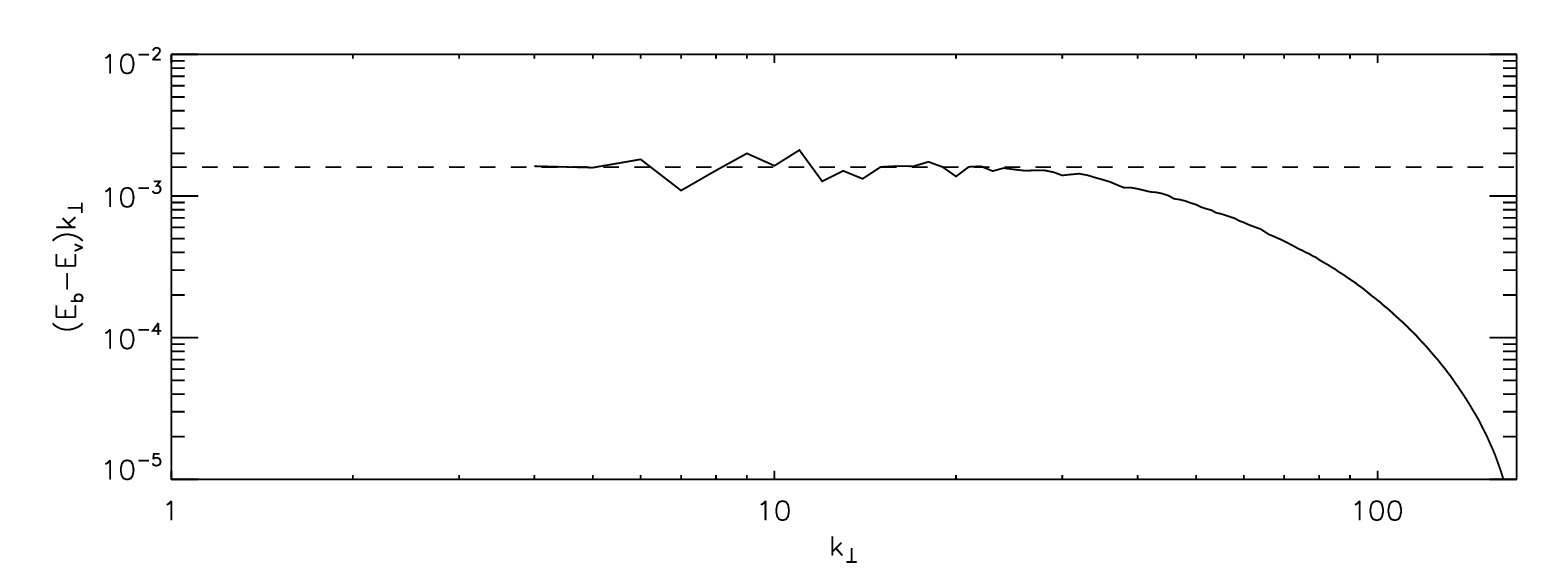}
\caption{Upper panel: The (normalized) residual energy $E_r(k_\|, k_\perp)/E_r(0, k_\perp)$ plotted as a function of $k_\|$ and $k_\perp$ for the weak balanced 
MHD turbulence. Lower panel: The field-perpendicular spectrum of the residual energy for weak balanced MHD turbulence compensated by $k_\perp$. At small $k_\perp$ the deviations are strong due to proximity to the forcing region (not shown). }
\label{fig:level}
\end{figure}

We find that the residual-energy condensate is indeed generated in weak balanced MHD turbulence, as shown in Fig.~(\ref{fig:cond}). The condensate is negative, and its broadening with $k_\perp$ shown in Fig.~(\ref{fig:level}) is consistent with the linear law predicted in Eq.~(\ref{wedge}). The scaling behavior of the field-perpendicular spectrum of the residual energy is also consistent with the analytic prediction (\ref{erperp}), see Fig.~(\ref{fig:level}). When we repeated the computation for the imbalanced case, that is, the case when $q^+\neq q^-$ (not presented here), we did not detect essential changes in the structure and scaling of the condensate. 

\section{Discussion}
In this Letter we demonstrated that residual energy is spontaneously generated by nonlinearly  interacting Alfv\'en waves, and it accumulates in a narrow region of phase space (\ref{wedge}). This explains the phenomenon of `condensate' of the residual energy numerically observed in \citep{boldyrev_p09}. Although the total (phase-space integrated) residual energy is small, its presence is crucial for the turbulent dynamics, for the following reason. Inside region (\ref{wedge}), that is, for $k_\|\approx 0$ the residual energy is comparable to the Elsasser energies and it follows the same  scaling, $e_r(0, k_\perp)\sim q^{\pm}(0, k_\perp) \propto k_\perp^{-3}$. Since the turbulent cascade of Alfv\'en waves crucially depends on the modes with $k_\| \approx 0$ \citep[e.g.,][]{montgomery_t81,sridhar_g94,ng_b96,galtier_nnp00,galtier_nnp02}, 
the residual energy crucially affects the dynamics of the Elsasser modes. 

More formally, this is expressed in the fact that one of the Elsasser energies enters the kinetic equations in the combination $q^{\pm}(k_\|, k_\perp){\tilde \delta}(k_\|)$, where ${\tilde \delta}(k_\|)$ are concentrated in the region (\ref{wedge}), see, e.g., Eq.~(\ref{bp-eq}). In previous treatments of weak MHD turbulence \citep{galtier_nnp00,galtier_nnp02}, it was assumed that $q^\pm(k_\|, k_\perp)$ are  smooth functions of $k_\|$, that is, their dynamically important components with $k_\| = 0$ have the same $k_\perp$-scaling as the  components with $k_\|\neq 0$. Such an assumption would be self-consistent if the residual energy were absent. However, as we have demonstrated, the residual energy is spontaneously generated by interacting Alfv\'en waves.  One can argue that the presence of the residual energy modifies the spectra of the Elsasser energies that, together with their smooth parts, $q^{\pm}=f^{\pm}(k_\|)k_\perp^{-3}$, now acquire their own singular parts  $\delta q^{\pm}= \alpha^{\pm}\epsilon^2 \Delta^{\pm}(k_\|)k_\perp^{-2}$,  where $\Delta^{\pm}(k_\|)$ are concentrated in region (\ref{wedge}). When multiplied by ${\tilde \delta}(k_\|)$ and integrated over $k_\|$, both the smooth and the singular parts provide comparable contributions to the integrals. 

We established that the spontaneously generated residual energy is always negative. Although this result is obtained in the framework of weak MHD turbulence, it provides the first analytic explanation for the observational and numerical findings that magnetic energy exceeds kinetic energy in the inertial interval of MHD turbulence, \citep[e.g.,][]{muller_g05,podesta_rg07,boldyrev_p09}.  

We derived that the residual energy has the field-perpendicular spectrum $E_r(k_\perp) \propto k_\perp^{-1}$ (\ref{erperp}). This relatively shallow spectrum holds in the inertial interval and breaks down at sufficiently large $k_\perp$ when the nonlinear broadening of the residual energy spectrum in the field-parallel direction (\ref{wedge}) becomes comparable to 
the width of the field-parallel energy spectra of the Alfv\'en waves, $q^{\pm}(k_\|)$. It is easy to see,  however, that this is precisely the scale beyond which the weak interaction approximation breaks down, and the turbulence becomes strong. In our future work we will extend our analysis of residual energy to the case of strong MHD turbulence. 

\acknowledgments
This work was supported   
by the US DoE Awards DE-FG02-07ER54932, DE-SC0003888, DE-SC0001794, the NSF Grant PHY-0903872, the NSF/DOE Grant AGS-1003451, and in part by the NSF Grant No. NSF PHY05-51164, and the NSF Center for Magnetic Self-organization in Laboratory and Astrophysical Plasmas at U. Wisconsin-Madison. High Performance Computing resources were
provided by the Texas Advanced Computing Center (TACC) at the
University of Texas at Austin under the NSF-Teragrid Project
TG-PHY080013N. 


\end{document}